**A gate-defined silicon quantum dot molecule**

Hongwu Liu [a)]

*NTT Basic Research Laboratories, NTT Corporation, 3-1 Morinosato-Wakamiya, Atsugi 243-0198, Japan; SORST-JST, Chiyoda, Tokyo 102-0075, Japan and National Laboratory of Superhard Materials, Institute of Atomic and Molecular Physics, Jilin University, Changchun 130012, P.R. China*

Toshimasa Fujisawa

*NTT Basic Research Laboratories, NTT Corporation, 3-1 Morinosato-Wakamiya, Atsugi, Japan and SORST-JST, Chiyoda, Tokyo 102-0075, Japan*

Hiroshi Inokawa,[b)] Yukinori Ono, and Akira Fujiwara

*NTT Basic Research Laboratories, NTT Corporation, 3-1 Morinosato-Wakamiya, Atsugi, Japan*

Yoshiro Hirayama

*Department of Physics, Tohoku University, Sendai, Miyagi 980-8578, Japan and SORST-JST, Chiyoda, Tokyo 102-0075, Japan*

We report electron transport measurements of a silicon double dot formed in multi-gated metal-oxide-semiconductor structures with a 15-nm-thick silicon-on-insulator layer. Tunable tunnel coupling enables us to observe an excitation spectrum in weakly coupled dots and an energy level anticrossing in strongly coupled ones. Such a quantum dot molecule with both charge and energy quantization provides the essential prerequisite for future implementation of silicon-based quantum computations.

[a)] Electronic mail: liuhw@ncspin.jst.go.jp

[b)] Present address: Research Institute of Electronics, Shizuoka University, 3-5-1 Jyouhoku, Hamamatsu, Japan

A gate-defined GaAs quantum dot molecule (QDM) with both charge and energy quantization has been applied to the implementation of charge and spin quantum bits (qubits) for solid-state quantum computing.[1-3] In such pioneering experiments, tunable gate barriers are required for precise control of quantized energy levels in the dots, the tunneling rate through a barrier, and the interaction of electron wave functions of the double dot (DD).

A QDM made from alternative materials such as silicon (Si) is another promising proposal for qubit operations because of the expected long spin relaxation and coherence time in Si.[4-6] So far, several efforts towards this goal have been initiated using DDs formed in metal-oxide-semiconductor (MOS) structures[7-10] and Ge/Si core/shell nanowires[11] as well as in Si/Ge two-dimensional electron gases[12]. However, clear evidence of fully tunable Si QDMs has not yet been reported due to a major technical difficulty: the size of each gate-defined Si dot needs to be small enough to achieve well-separated energy levels since the effective mass in Si is relatively large. Moreover, in a MOS structure, parasitic dots are easily formed due to either the oxidation process in Si nanostructures or the roughness at the $Si/SiO_2$ interface.[13,14] More recently, we have fabricated a Si DD in multi-gated MOS structures with a 10-nm-thick Si-on-insulator (SOI) nanowire.[15] In such a device, the roughness at the Si/buried oxide interface induces significant potential fluctuations in the thin SOI layer and thus parasitic dots are readily formed in the transport channel. Two neighboring gates occasionally enable us to confine a quite small Si DD, in which quantized energy levels are well separated. However, there is no independent electrical gate to tune the coupling between the two dots.

In this letter, we present a Si DD formed in a thick (15 nm) SOI layer. The relatively thick SOI layer is expected to tame the significant potential fluctuations and allows us to obtain a fully tunable Si QDM. In a weakly coupled DD, we have observed resonant tunneling through quantized energy levels, which is dominated by electrostatic (capacitive) interdot coupling. In a strongly coupled DD, an energy level anticrossing induced by quantum interdot tunneling is clearly presented. Since one can take advantage of mature MOS technologies to integrate large numbers of qubits, our experiment represents an important step toward Si-based qubit operations.

A schematic cross section of the multi-gated MOS structure is shown in Fig. 1(a).[16] By applying positive voltage $V_{UG}$ to the upper poly-Si gate (UG), an electron inversion layer is created between the gate oxide and the SOI layer. Negative voltages applied to the three lower poly-Si gates (LG1, LGC and LG2) repel electrons underneath to form tunnel barriers of a DD. All experiments were performed using dc measurements in a dilution refrigerator with a base temperature of 30 mK at zero magnetic field. Measurements of Coulomb diamonds of single dots gave a charging energy $E_C \sim 7$ meV (with a conversion factor $\alpha_l$ of 0.4 eV/V relating the lower-gate voltage to the bias potential) and a quantized level spacing $\Delta E \sim 0.5$ meV. The total capacitance $C$ of the single dot ($\sim 25$ aF) is calculated from $C = e^2/(E_C-\Delta E)$. The dot dimension of $65\times50$ nm$^2$ is estimated by solving the capacitance of an elliptic conducting plate of our dot.[17] Considering a twofold degeneracy of spin and the lifting of the valley degeneracy,[15] we estimated the level spacing[18] in such a dot to be 0.4 meV. This estimation is comparable to the measured $\Delta E$. The number of electrons in the dot evaluated from an areal electron density of $3\times10^{12}$ cm$^{-2}$ in the inversion layer was about 80. An analysis of temperature-dependent linewidths of linear conductance peaks gave an electron temperature $T_e \sim 0.8$ K. These results show clear evidence of a single-level transport in single dots with $E_C > \Delta E > k_B T_e$ ($k_B$ is Boltzmann's constant).[19]

Current $I$ through the DD at gate voltage $V_{LGC} = -630$ mV and source-drain voltage $V_{SD} = 1$ mV is shown in Fig. 1(b), where a hexagonal charge domain is marked by solid lines. Charge states (n,m) denote effective electron numbers in the left (n) and right (m) dots tuned by gate voltage $V_{LG1}$ and $V_{LG2}$ respectively. Peaks (● and ○, also named triple points) represent single-electron and single-hole sequential tunneling through the DD and evolve into triangles at finite $V_{SD}$.[20] The peak separation $\Delta V_S$ is found to be dependent on interdot coupling tuned via $V_{LGC}$. Figure 1(c) shows the measured fractional splitting $F = 2\Delta V_S /\Delta V_P$[21,22] versus $V_{LGC}$, where $\Delta V_P$ is the period of charge domains along a diagonal of the $V_{LG1}$-$V_{LG2}$ plane [Fig. 1(b)]. The F (or $\Delta V_S$) increases from nearly zero to one (or $\Delta V_P/2$) with decreasing $|V_{LGC}|$ (corresponding to an increase in interdot coupling), indicating a flexible control of the device from a completely decoupled DD to a merged single dot. Similar charge modulations have also been reported in another sample with a 20-nm-thick SOI layer.[8]

In such a tunable interdot coupling regime, there exist two interaction mechanisms: one is the capacitive coupling corresponding to the change in the electrostatic potential of one dot when that of another changes, and the other is the quantum tunneling due to a direct coupling of electron wave functions in the two dots. The honeycomb charge domain in Fig. 1(b) indicates the dominant electrostatic interdot coupling in this DD. All capacitances that characterize the DD can be calculated from the size of the triangles and the hexagons. The total capacitance $C_{1(2)} \sim 23(20)$ aF with respect to the left (right) dot yields a charging energy $E_C \sim 7$ (8) meV, which is consistent with measurements on single dots. The mutual capacitance $C_m$ between the two dots is 2 aF, corresponding to an electrostatic coupling energy $E_{cm} \sim 0.7$ meV. Figure 2(a) shows a fine scan of one pair of triangles measured at $V_{LGC}$ = -660 mV (marked by solid lines). The bases of the triangles along two triple points represent resonant tunneling through the ground states of the DD, while lines parallel to the bases manifest excited-state resonances. The ground-state resonance shows weak current, probably arising from a bad coupling of the ground states to the leads. By sweeping gate voltage along the arrow ε, the current as a function of detuning energy ε between discrete levels of the DD was obtained, as plotted in Fig. 2(b). Four resonant peaks are presented, but they are not well separated because of broad linewidths. For simplicity, we neglect the spontaneous emission contribution and fit the data using four thermally broadened resonances $I(\varepsilon) \sim \cosh^{-2}(\varepsilon/w k_B T_e)$ with $w \approx 2.75$ and $T_e \sim 0.8$ K (dashed lines). The fit, as shown by a solid gray line, is found to coincide with the data, indicating that the thermal broadening is the origin of the linewidths. The fitting parameter $T_e$ is the same as the measured value in single dots. Note that in our case the $T_e$ primarily originates from electrical noise in the measurement systems and is expected to be further filtered using a combination cryogenic filter.[23] Lowering $T_e$ would result in a more distinct excitation spectrum with narrow resonance lines. In addition, as shown in Fig. 2(b), the average peak spacing of 0.3 meV is comparable to the $\Delta E$ in single dots. Figure 2(c) shows the current $I_{res}$ and the full width at half-maximum (FWHM) of the excited-state peak (◇) as a function of $V_{LGC}$. The $I_{res}$ increases with decreasing $|V_{LGC}|$, suggesting that the electrostatic interdot coupling is smaller than the tunnel coupling of incoming and outgoing barriers and thus dominates the current.[24,25] The FWHM of the resonance is independent of $V_{LGC}$ because it is determined by the thermal broadening. This is in contrast to intrinsic resonant tunneling through DDs, in which the FWHM strongly depends on the interdot coupling.[25] Similar results were obtained from other resonances of the DD. Thus, we have demonstrated excitation spectra of the weakly coupled Si DD, in which resonant tunneling current is dominated by tunable capacitive interdot coupling.

Further decreasing $|V_{LGC}|$ allows us to investigate the DD in a strong coupling regime. Figure 3(a) shows the current $I$ through the DD at $V_{LGC}$ = -570 mV and $V_{SD}$ = 30 μV. As is clearly seen, two triple points (● and ○) evolve into a level anticrossing, indicating the dominant quantum interdot tunneling ($t$) in this DD. The $t$ can be derived from the plot of the energy separation $\Delta E$ of the anticrossing versus ε as shown in Fig. 3(b). One can see that the solid line calculated by $\Delta E = E_{cm} + \sqrt{\varepsilon^2 + 4t^2}$ with a parameter of $t$ ~ 0.4 meV is consistent with the data. Figure 3(c) shows the dependence of $t$ on $V_{LGC}$. The obtained gate-tunable $t$ makes it possible to swap two electron spins of the DD by applying a pulsed $V_{LGC}$. Taking the Heitler-London model into account,[26] we determined the operation time $\tau_{swap} = \pi\hbar(E_C - E_{cm})/4t^2$ ($\hbar$ is the reduced Plank's constant) to be less than 80 ps [Fig. 3(c)]. The swap time is much shorter than the expected decoherence time (a few ten milliseconds),[6] suggesting the suitability of our Si DDs for qubit operations.

In conclusion, we took advantage of the MOS technique to produce a fully tunable Si QDM. The weakly coupled DD with resolved quantum energy levels would be useful for investigating elastic and inelastic tunneling and elucidating the electronic states with respect to the interplay between the orbit, valley, and spin degrees of freedom in Si DDs.[27] The strongly coupled DD with anticrossing levels makes it possible to probe the nature of single-electron charge and spin coherence in Si nanodevices by performing pulse measurements. Our Si QDM with a fast quantum tunneling rate is feasible for producing fundamental quantum gates for spin qubit operations with the expected long spin coherence time.

This work was partially supported by JSPS KAKENHI (16206038 and 19310093). H. W. Liu thanks the National Natural Science Foundation of China under Grant No. 10574055 and the Program for New Century Excellent Talents in University.

FIG. 1. (color online) (a) Schematic cross section of our Si MOS device. Three lower gates, LG1, LGC, and LG2, were used to form tunnel barriers for the dots. The widths of the lower gate and the SOI nanowire are 65 and 55 nm. The gate spacing is 85 nm. Thicknesses of the buried oxide, the gate oxide, and the oxide around the lower gate are 400, 35, and 30 nm, respectively. (b) Current $I$ flowing through the DD as a function of $V_{LG1}$ and $V_{LG2}$ at $V_{LGC}$ = -630 mV and $V_{SD}$ = 1 mV. (c) Measured fractional splitting F = $2\Delta V_S /\Delta V_P$ versus $V_{LGC}$.

FIG. 2. (color online) (a) One pair of triangular conductive regions in Fig. 1(b) measured at $V_{LGC}$ = -660 mV and $V_{SD}$ = 1 mV. (b) Current (solid black line) plotted along the arrow ε in (a). The solid gray line is a fit with four thermal-broadened peaks shown by dashed gray curves. (c) Current $I_{res}$ (■) and FWHM (★) of the excited-state resonance [◇ in (a) and (b)] as a function of $V_{LGC}$. Dotted lines are guide to the eyes.

FIG. 3. (color online) (a) Current $I$ through a strongly coupled DD as a function of $V_{LG1}$ and $V_{LG2}$ at $V_{LGC}$ = -570 mV and $V_{SD}$ = 30 μV. The $E_{cm} \approx$ 1 meV is estimated at $V_{LGC}$ = -610 mV. The bars ΔE (separation energy of the anticrossing) and ε (detuning energy between quantized levels of the DD) denote the new axis directions. The coordinate origin is at the center of the dashed line connecting two triple points (● and ○). Nonlinear transport at large $V_{SD}$ determines a conversion factor $\alpha_2 \sim$ 0.2 eV/V relating the gate voltage to the electrochemical potential, $\Delta E (\varepsilon) = \alpha_2 V_{\Delta E(\varepsilon)}$. (b) Plot of ΔE as a function of ε. The solid curve is a fit to the data. (c) Dependence of quantum tunneling rate $t$ and swap-gate operation time $\tau_{swap}$ on $V_{LGC}$.

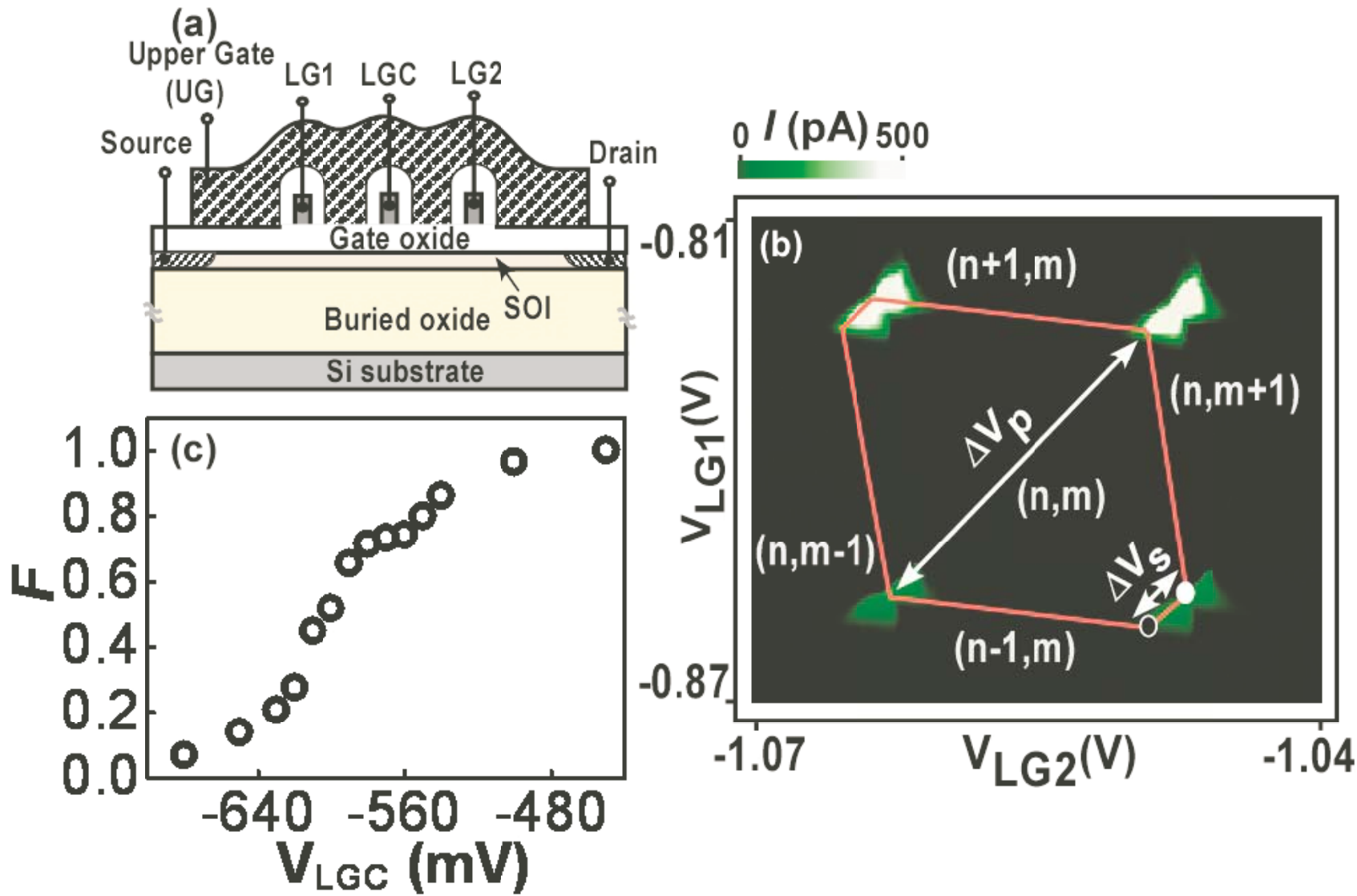

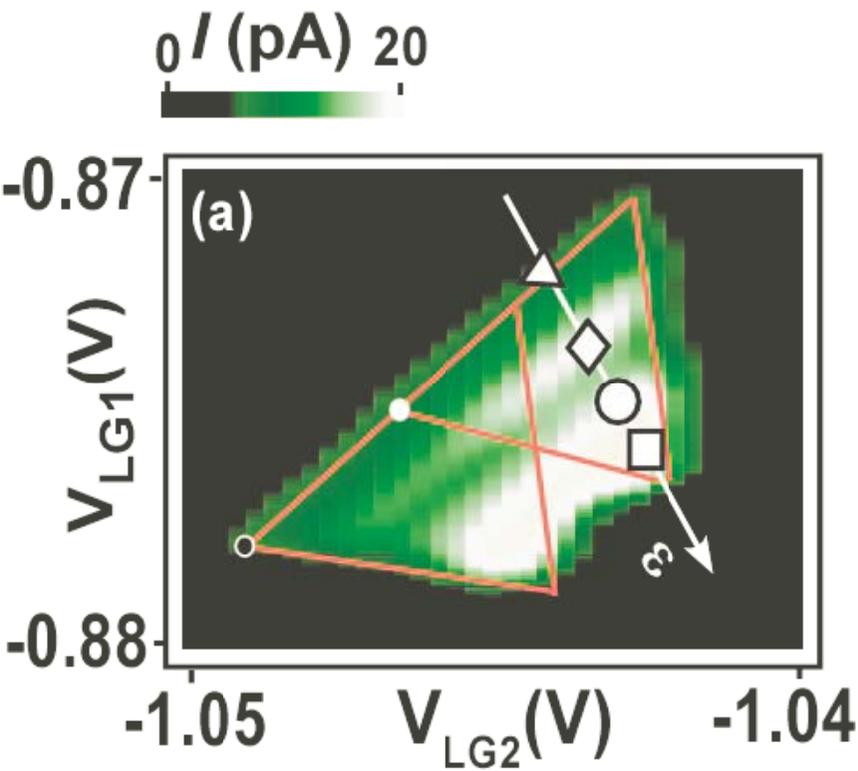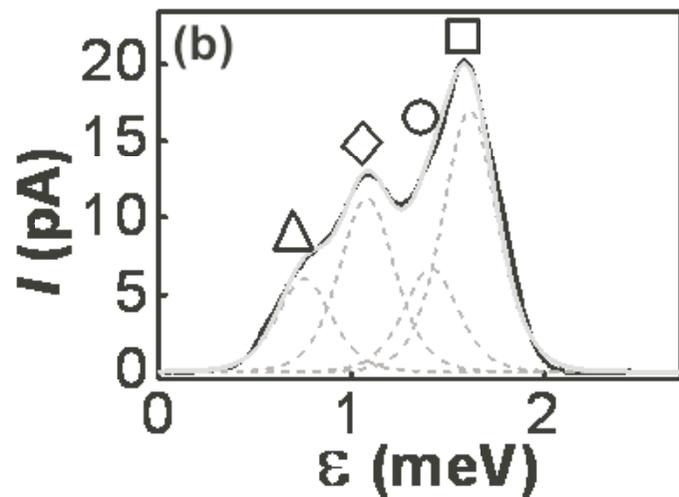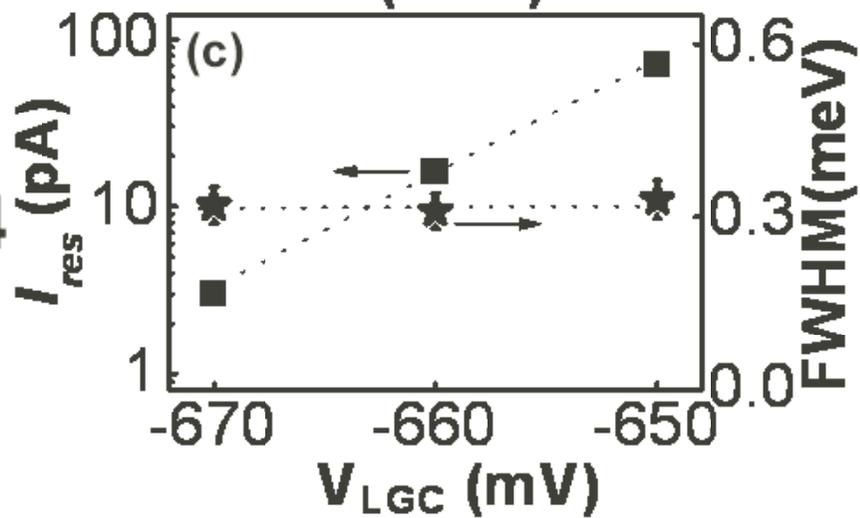

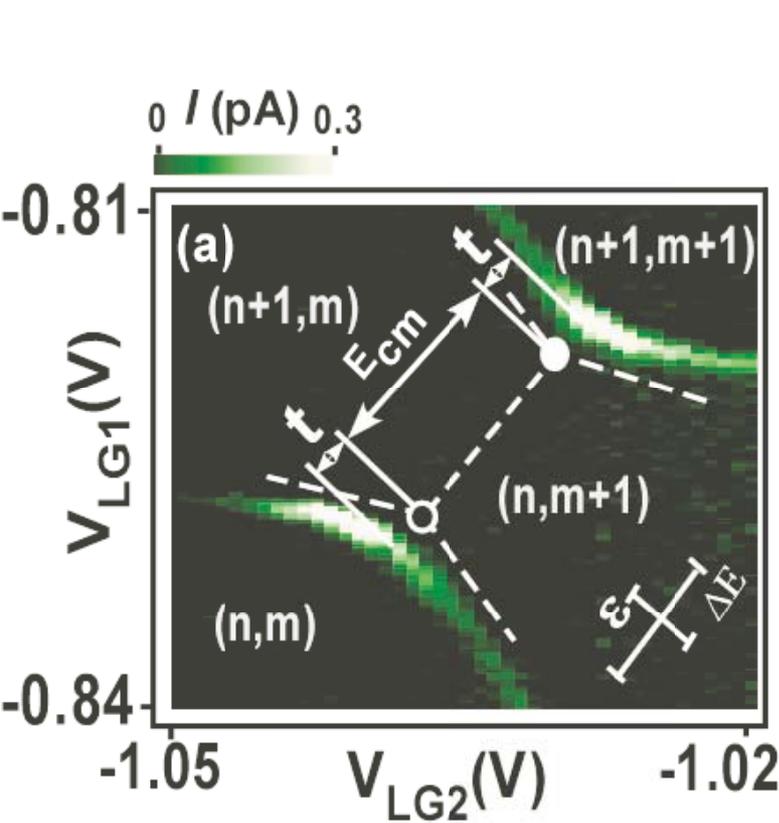
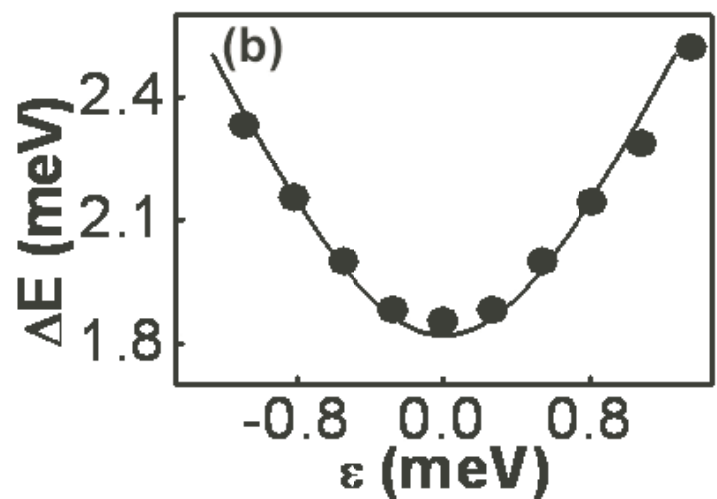
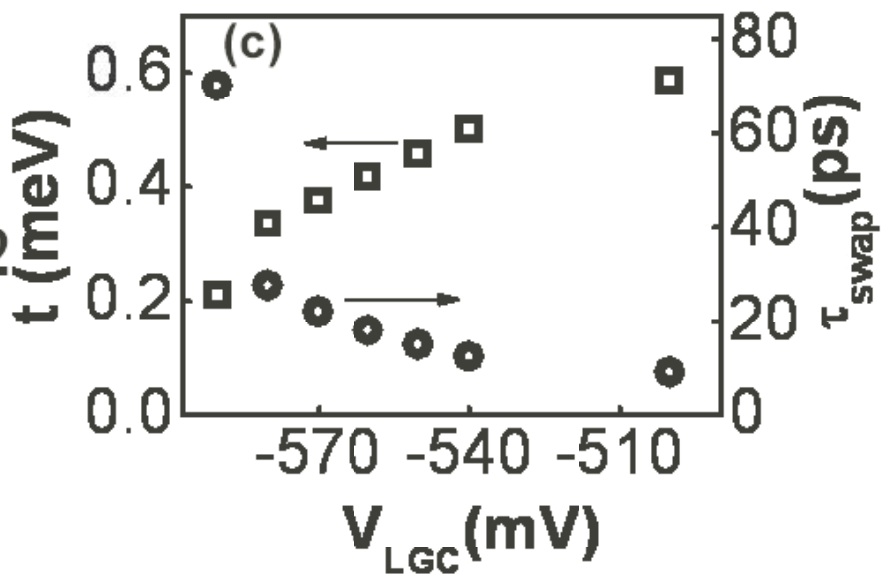